\documentstyle[12pt,aasms4]{article}

\lefthead{Bekki Kenji}
\righthead{The Fundamental Plane and merger scenario}

\begin{document}
\title{The Fundamental Plane and merger scenario I.
  Star formation
history of galaxy mergers  and  origin of the Fundamental Plane}

\author{Kenji Bekki} 
\affil{Astronomical Institute, Tohoku University, Sendai, 980-77, Japan}

\begin{abstract}

 We perform  numerical simulations of  galaxy mergers between star-forming
and 
gas-rich spirals 
in order 
to explore the origin of the Fundamental Plane (FP)
of elliptical galaxies.
We consider particularly  that the origin of the slope of the FP  is essentially
due to the non-homology in structure and kinematics of  elliptical galaxies
and  accordingly
investigate structural and kinematical properties of 
elliptical galaxies formed by dissipative galaxy merging with
star formation. 
We found that the rapidity of star formation, which is defined as the ratio
of dynamical time-scale of merger progenitor 
to the time-scale of gas consumption by  star formation, 
is a key determinant for nonhomology parameters, 
such as the density profile of stellar component, the relative importance
of global rotation in kinematics,
and the ratio of total dynamical mass to luminous mass, in merger remnants.
We furthermore found that this result
does not depend so strongly on initial intrinsic spins of progenitor disks and
orbital energy and angular momentum of mergers. 
These  results strongly  suggest that the structural and kinematical
nonhomology observed in elliptical galaxies
can be  closely associated with the difference in  star formation history 
between elliptical galaxies formed by dissipative galaxy merging. 
Based upon these results, we discuss a  close physical relation between
the origin of the FP and the star formation history of elliptical galaxies.
 
\end{abstract}


\section{Introduction}

 The Fundamental Plane (FP) is one of the most important universal relations
 in early type galaxies because it contains valuable information about the
 formative and evolutionary process  of galaxies
 (\cite{dd87,dre87}).
 The commonly used form of the scaling relation in the FP is described as
 $R_{\rm e}$=
 $\sigma^A I^B$,
 where $R_{\rm e}$, $\sigma$, and $I$
 are effective radius, central
velocity dispersion, and mean surface brightness of
 elliptical galaxies,
 respectively.
 The exponents $A$, $B$ are considered to be $1.56 \pm 0.07$
 and $-0.94 \pm 0.09$ in the FP derived by $K$ band photometry,
 respectively,
 and these values  deviate significantly
 from the  values $A = 2.0$ and $B = -1.0$ expected
 from virial theorem (e.g., \cite{pah95}).
 This apparent deviation requires that the ratio of dynamical mass ($M$) to
 luminosity of elliptical galaxies  ($L$)  depends on $M$ 
as $M/L$
 $\propto$ $M^{\alpha}$ ($\alpha$ = 0.12 $\pm$ 0.03 for $K$ band).
 To understand the physical meaning of this dependence is one of the
 fundamental issues in the theory of galaxy formation (e.g,
Bender, Burstein, \& Faber 1992, 1993).
Possible interpretations for the required dependence of $M/L$ on $M$
(the ``slope'' or ``tilt'' of the FP)
are generally considered to be divided into the following two.
One is that the required dependence of  $M/L$ on $M$ 
results from the fact that
the mean stellar age and metalicity of elliptical galaxies
depend systematically  on  $M$.
The other  
is that the required dependence  reflects the 
 $M$  dependence of   structural
and kinematical properties of elliptical galaxies (``nonhomology''). 
Although we should not neglect the importance of stellar populations
in generating the $M$  dependence  of the $M/L$ 
(Renzini \& Ciotti 1993;
Pahre \& Djorgovski 1997), we here consider that
the origin of the  required $M$   dependence of $M/L$ 
is closely associated with the structural
and kinematical properties dependent on $M$ or $L$
in
elliptical galaxies.

A growing number of observational studies have been accumulated which reveal the 
nonhomology in structural and kinematical properties of elliptical galaxies,
such as the  systematical departure from the universal light profile of
elliptical galaxies,
the so-called $R^{1/4}$ law (e.g., Burkert 1993; Caon, Capaccioli,
\& D'Onofrio 1993), the difference in the relative importance of
global rotation in kinematics between elliptical galaxies (e.g., Davies et al. 
1983), and the ratio of dynamical mass to luminous mass (e.g., Rix 1997). 
These nonhomology observed in  elliptical galaxies are considered to be one of
the most promising candidates which can provide  a plausible explanation 
for the origin of the slope of the FP (Ciotti, Lanzoni,
\& Renzini 1996; Pahre et al. 1995;
Prugniel \& Simien 1996; Graham \& Mathew
1997).
There are a number of  theoretical attempts to provide  a physical basis 
for the origin
of the above structural and kinematical nonhomology of elliptical galaxies.
Hjorth \& Madsen (1995) constructed a self-consistent equilibrium model
of elliptical galaxies based upon the statistical
mechanics of violent relaxation and demonstrated that if the more luminous 
elliptical galaxies have the so-called  $R^{1/n}$ light profile with larger $n$, the
slope of the FP can be successfully reproduced.
Capelato, de Carvalho, \& Carlberg  (1995) 
performed  a set of numerical simulations on  dissipationless
galaxy mergers with variously different initial 
conditions of mergers and found that initial orbital energy of merging systems is 
a primarily important factor for the radial profile of
velocity dispersion and mass distribution of merger remnants
and thus for the successful reproduction of  the FP slope. 
These studies have
indeed provided clues about  the understanding of the origin of the FP slope,
it has not been clarified,  however,  why and how 
the  exponent  $n$ in the $R^{1/n}$ light profile
and the initial orbital energy of merging systems 
depend primarily on the galactic luminosity or mass. 
Thus it is vital to make it clear why and how the structural and kinematical
nonhomology required for the reproduction of the FP slope depends on
the galactic luminosity or mass.

The purpose of this paper is to elucidate the origin of the 
luminosity-dependent structural and kinematical properties
(``nonhomology'') in elliptical galaxies
and thereby to explore the physical meaning of the FP.
We here focus particularly on the star formation
history of elliptical galaxies  and accordingly
investigate the important roles of the star formation history 
in generating the structural and kinematical nonhomology
of elliptical galaxies.
In the present study, the origin of the  luminosity-dependent structural 
and kinematical properties (nonhomology) is considered as 
follows.
First, the difference in structural and kinematical properties of elliptical galaxies
is determined primarily by the difference in the star formation history of
elliptical galaxies.
Next, the star formation history of forming elliptical galaxies
depends  principally on galactic luminosity.
The structural and kinematical properties of elliptical galaxies 
therefore depend on galactic luminosity. 
This consideration for the understanding of the origin of the FP slope
is reasonable and realistic,
 since difference in 
 star formation history between elliptical galaxies
 and its possible dependence on galactic luminosity
 are suggested both by theoretical studies and by
 observational ones  on  chemical and photometric evolution
 of elliptical galaxies such as  color-magnitude relation
(e.g., \cite{ay87}),
 the abundance ratio [Mg/Fe] 
 (\cite{wor92,mat94,ben96}),
and $H_{\beta}$ line index (Faber et al. 1995; Worthey, Trager, \& Faber 1996).
In investigating the non-homologous nature in elliptical galaxies,
we adopt the ``merger hypothesis'' in which elliptical galaxies are
formed by galaxy mergers between two late-type spirals (Toomre \& Toomre 1977).  
 In particular, we consider galaxy mergers between two disk galaxies
with a gas mass
fraction larger than 0.2 observed in the present typical late-type
spirals 
in order to mimic elliptical galaxy formation
by galaxy mergers at higher redshift,
which is suggested by 
recent observational studies on the tightness of the color-magnitude relation
and the FP and on the 
photometric evolution of elliptical galaxies
(Bower et al. 1992;
Ellis et al. 1996; Franx \& van Dokkum 1996).
We do not here intend to discuss  formation of elliptical
galaxies within a framework of a specific cosmology such as the
CDM model (e.g., White \& Rees 1978; White \& Frenk 1992; Baugh, Cole,
\& Frenk 1996).

The  layout of this paper is as follows.
In \S 2, we summarize numerical models used in the present study and main points
of analysis of the study.
In \S 3, we describe the dependence of  structural and kinematical properties
of merger remnants on the star formation history of galaxy mergers.
In this \S 3, we furthermore discuss how
the  star formation history of galaxy
mergers should depend on galactic luminosity
for the successful reproduction of the FP slope observed 
in elliptical galaxies.
In \S 4, we discuss a close relation between the origin of the slope
and the tightness of the FP and the star formation history of galaxy mergers.
The conclusions of the present study are given in \S 5.

\section{Model}

\subsection{Numerical model}

\subsubsection{Initial conditions}

 We construct  models of galaxy mergers between gas-rich 
 disk galaxies with equal mass by using Fall-Efstathiou model (1980).
 The total mass and the size of a progenitor disk are $M_{\rm d}$
 and $R_{\rm d}$, respectively. 
 From now on, all the mass and length are measured in units of
  $M_{\rm d}$ and  $R_{\rm d}$, respectively, unless specified. 
  Velocity and time are 
  measured in units of $v$ = $ (GM_{\rm d}/R_{\rm d})^{1/2}$ and
  $t_{\rm dyn}$ = $(R_{\rm d}^{3}/GM_{\rm d})^{1/2}$, respectively,
  where $G$ is the gravitational constant and assumed to be 1.0
  in the present study. 
  If we adopt $M_{\rm d}$ = 6.0 $\times$ $10^{10}$ $ \rm M_{\odot}$ and
  $R_{\rm d}$ = 17.5 kpc as a fiducial value, then $v$ = 1.21 $\times$
  $10^{2}$ km/s  and  $t_{\rm dyn}$ = 1.41 $\times$ $10^{8}$ yr,
  respectively.
  In the present model, the rotation curve becomes nearly flat
  at  0.35  radius with the maximum rotational velocity $V_{\rm max}$.
  In the present study,
this  $V_{\rm max}$ is assumed to be free parameter which controls
  the degree of self-gravitation of initial galactic disks.
  The  value of $V_{\rm max}$ for each model is described later. 
  If we assume, for example,
 that $V_{\rm max}$ is equal to  1.8 which is used for
  most of models in the present study, 
  the corresponding total mass $M_{\rm t}$ and halo mass $M_{\rm h}$
  are 3.8 and 2.8 in our units, respectively.
  The radial ($R$) and vertical ($Z$) density profile 
  of a  disk are  assumed to be
  proportional to $\exp (R/R_{0}) $ with scale length $R_{0}$ = 0.2
  and to  ${\rm sech}^2 (Z/Z_{0})$ with scale length $Z_{0}$ = 0.04
  in our units,
  respectively.
  The mass density  of halo component is truncated at  1.2 in our units
  and its velocity dispersion at a given point
  is set to be isotropic and given
  according to the  virial theorem.
  In addition to the rotational velocity made by the gravitational
  field of disk and halo component, the initial radial and azimuthal velocity
  dispersion are given to disk component according
  to the epicyclic theory with Toomre's parameter (\cite{bt87}) $Q$ = 1.0.
  This adopted value for $Q$ parameter
is appreciably smaller compared with the value required 
 for stabilizing the initial disk
against the non-axisymmetric dynamical instability (e.g. bar instability).
The reason for this adoption is that the initial disk is assumed to
be composed mostly of interstellar gas and thus random kinetic energy
in the disk is considered to be rather smaller because of gaseous
dissipation in  the disk. 
  The vertical velocity dispersion at given radius 
  are set to be 0.5 times as large as
  the radial velocity dispersion at that point, 
  as is consistent  with 
  the observed trend  of the Milky Way (e.g., Wielen 1977).
Actually, the initial structure and kinematics in precursor disks of a merger
are   different between galaxy mergers  and depends on the epoch of
the merging.
Although this difference probably can yield a great variety of 
dynamical structures in merger remnants, we do not intend to
consider this important difference for simplicity in the present  paper
and will address in our future paper.

  The collisional and dissipative nature 
  of  interstellar medium is  modeled by the sticky particle method
  (\cite{sch81}).
It should be emphasized here that this discrete cloud model can at best represent
the $real$ interstellar medium of galaxies  in a schematic way. 
As is modeled by McKee \& Ostriker (1977),
the interstellar medium can be considered to be  
 composed mainly of `hot', `warm', and `cool'
gas,
each of which mutually
interacts hydrodynamically 
 in a rather  complicated way.
 Actually, these considerably complicated nature of
interstellar medium in  disk galaxies would not be
  so simply modeled by the `sticky
particle' method in which gaseous dissipation is modeled by ad hoc
cloud-cloud collision: Any existing numerical method probably could
not model the $real$ interstellar medium in an admittedly proper
way. 
In the present study, as a compromise,
we only try to address some important aspects of hydrodynamical
interaction between interstellar medium in disk galaxies and in
dissipative mergers. 
More elaborated numerical modeling for real interstellar medium
would be  necessary for 
our further understanding of dynamical evolution 
in dissipative galaxy mergers. 
  To mimic the
  galaxy mergers which are 
occurred at   higher redshift and thus very dissipative
because of a  considerably larger amount of  interstellar gas in the 
progenitor disks,
  we assume that the fraction of gas mass 
  in a disk, which is  represented by
 $M_{\rm g}$, is  a free parameter with the value ranging from 0.1
 to 1.0.
The value of  $M_{\rm g}$ for each model is described later. 
  The size  of the clouds in the sticky particle method
is set to be 3.5 $\times$ $10^{-3}$ in our units 
  in the present simulations. 
  The radial and tangential restitution coefficient for cloud-cloud
  collisions are
  set to be 0.5 and
  0.0, respectively.

    In all of the simulations, the orbit of the two disks in a
merger is set to be
    initially in the $xy$ plane and the distance between
    the center of mass of the two disks,
represented by $r_{\rm in}$,
is  assumed  to be the  free parameter 
which controls the epoch of galaxy merging.
    The pericenter
    distance, represented by $r_{\rm p}$, is also
assumed to be the  free parameter which controls the initial
total orbital  angular momentum of galaxy mergers.
The orbit eccentricity, represented by $e_{\rm p}$,
is  also assumed to be the free parameter which
controls initial orbital energy of galaxy mergers.
    The spin of each galaxy in a  merger
is specified by two angle $\theta_{i}$ and
    $\phi_{i}$, where suffix  $i$ is used to identify each galaxy.
    $\theta_{i}$ is the angle between the $z$ axis and the vector of
    the angular momentum of a disk.
    $\phi_{i}$ is the azimuthal angle measured from $x$ axis to
     the projection of the angular momentum vector of a disk on
    to $xy$ plane. 
The time when the progenitor disks merge completely and reach  the
dynamical equilibrium is less than 28.0  in our units for most of
models and does not depend so
strongly on the  history of star formation in the  present calculations.
The value of each parameter,  $\theta_{i}$, $\phi_{i}$, $r_{\rm p}$ 
$r_{\rm in}$, and $e_{\rm p}$ for each model is described later.

As is described above, the present merger model is different  from that of previous
studies in the following two points.
Firstly, the initial gas mass fraction is assumed to be 1.0 for most of models in the
present study, which is considerably different from 0.1 (0.2) adopted in previous studies.
This is principally because we intend to investigate dynamical evolution of relatively
higher redshift galaxy mergers, which probably have a larger amount of interstellar gas.
Secondly, the radius at which the rotation curve of an initial disk becomes flat is assumed
to be 
relatively larger, which is one of characteristics of late-type spiral galaxies.
Considering our not including the bulge component in merger progenitor galaxies,
the present  model describes galaxy mergers between `purely' late-type disks 
considerably abundant in interstellar gas.

\subsubsection{Global star formation}

    Star formation
     is modeled by converting  the collisional
    gas particles
    into  collisionless new stellar particles according to the algorithm
    of star formation  described below.
    We adopt the Schmidt law (Schmidt 1959)
    with exponent $\gamma$ (1.0  $ < $  $\gamma$
      $ < $ 2.0, \cite{ken89}) as the controlling
    parameter of the rate of star formation.
 The  $\gamma$ is set to be 2.0 for all the simulations in
the present study.
    The amount of gas 
    consumed by star formation for each gas particle
    in each time step, 
    $\dot{M_{\rm g}}$, 
is given as:
    \begin{equation}
      \dot{M_{\rm g}} \propto  C_{\rm SF} \times 
 {(\rho_{\rm g}/{\rho_{0}})}^{\gamma - 1.0}
    \end{equation}
    where $\rho_{\rm g}$ and $\rho_{0}$
    are the gas density around each gas particle and
    the mean gas density at 0.48 radius  of 
    an initial disk, respectively.
    In order to avoid a large number of new stellar particles with
different mass, we convert one gas particle into one stellar one
according to the following procedure.
First we give each gas particle the probability, $P_{\rm sf}$,
 that the gas particle
is converted into stellar one, by setting the $P_{\rm sf}$ to be proportional
to the  $\dot{M_{\rm g}}$  in equation (1) estimated for the gas particle. 
Then we draw the random number to determine whether or not the gas particle
is totally converted  into one new star. 
This method of star formation enables us to control the rapidity of star formation
without increase of particle number in each simulation thus to maintain
the numerical accuracy in  each simulation. 
    The $C_{\rm SF}$ in the equation (1)
is the parameter that controls the rapidity of 
    gas consumption by star formation:
    The larger the $C_{\rm SF}$ is, the more rapidly the gas particles 
    are converted to new stellar particles.
As a result of this, total amount of gaseous dissipation is 
also controlled by this parameter $C_{\rm SF}$:
The more amount of kinetic energy of gas particles is dissipated away by
cloud-cloud collision for the models with smaller $C_{\rm SF}$.
This parameter $C_{\rm SF}$ is meant to be  proportional to
the ratio of dynamical time-scale of the system to the time-scale of
gas consumption  by star formation.
Furthermore the equation (1) states that
a larger number of stellar particles are created at the
regions  where the local gas density become larger owing to the onset
of local Jeans instability.
The positions and velocity of the new stellar particles are set to 
be the same as those of original gas particles.
As is described above, in the present study, we do not explicitly
include the `threshold density' of star formation,
which is demonstrated to be associated with the growth
of local gravitational
instability with  relatively smaller wavelength, that is, the Toomre's 
$Q$ parameter (Kennicutt 1989), for isolated disk galaxies.
This is because
the threshold criterion, which is derived only for  calmly evolving
galactic disks,
would not so simply be applied to the present merger model,
 in which 
the time-scale that the disk can evolve without strong dynamical 
perturbation is relatively smaller (less than  10 dynamical time for typical models).
 In  the present study, we do not intend to include any `feedback
effects' of star formation such as 
thermal and dynamical heating of interstellar medium driven by
supernovae, firstly because such  inclusion probably 
prevents us from deducing more clearly the important roles of the rapidity
of star formation and secondly because there still remains a great
uncertainty concerning the numerical  implementation of the `feedback
effects' (Katz 1992; Navarro \& White 1993).


\subsubsection{Parameter values for each model}
We perform  14 sets of numerical simulations of galaxy mergers
with variously different initial conditions of galaxy merging.
For most of models in the present study, the initial gas mass fraction
($M_{\rm g}$) is set to be 1.0, principally  because we intend to
elucidate more clearly the roles of star formation  in determining
structure and kinematics of merger remnants.
Each set of merger models is labeled
according to  initial
orbit configuration, gas mass fraction, and the maximum
value of rotation curve in galaxy mergers (Model 1 $\sim$ 14). 
The value of each parameter in  each set of merger models
is  summarized in Table 1. 
The first, second,  third , and  fourth
column denote the model number (Model
1 $\sim$ 14), maximum rotational velocity of a disk ($V_{\rm max}$),
the corresponding total mass in a merger progenitor ($M_{\rm t}$)
and initial gas mass fraction ($M_{\rm g}$), respectively.
The fifth,  sixth, and seventh  column give the value of initial
separation of galaxy mergers ($r_{\rm in}$), 
pericenter distance ($r_{\rm p}$),
and orbit eccentricity 
($e_{\rm p}$), respectively.
The values  of initial intrinsic spins of two disks,
$\theta_{i}$ ($i$ = 1, 2),  are given in the
eighth and ninth columns,
and those of $\phi_{i}$ ($i$ = 1, 2) in the  tenth and eleventh columns. 
Each set of models, Model 1 $\sim$ 14,
includes 
five $dissipative$ merger models with different rapidity of star formation
($C_{\rm SF}$ = 0.35, 0.7, 1.75, 3.5, and 7.0).
Model 1 is considered  to be the 
standard set of models  and the results of the set of models
are presented the most frequently in the present study.
Thus we  examine totally  70 merger models with variously different
initial conditions of galaxy merging in the present study. 
Although the adopted value of each parameter still would  not be so extended
enough to describe reasonably well the $real$ parameter space of galaxy merging
as is predicted by a specific cosmological model,
we however believe that the present parameter study can enable us to
provide clues about the origin of the FP.

   All the calculations related to 
dynamical evolution  of galaxy mergers including the dissipative
dynamics, star formation, and gravitational interaction between collisionless
and collisional component 
have been carried out on the GRAPE board
(\cite{sug90})
 at Astronomical Institute of Tohoku University.
The  number of halo particles and that of the gaseous ones
for an  isolated disk galaxy with $M_{\rm g}$ = 1.0 are
5000 and 10000, respectively.
For an isolated disk galaxy  with $M_{\rm g}$ = 0.1 and 0.5 
(for models in  Model 12 and 13),
the  number of particles of halo, stars, 
and gas 
are 5000, 5000, and 10000, respectively.
The parameter of gravitational softening length
is set to be fixed at 0.04  
for  models with $M_{\rm g}$ = 1.0
and 0.036 for models
with $M_{\rm g}$ = 0.1 and 0.5.
The reason for the smaller value of the  gravitational softening length
in the models with $M_{\rm g}$ = 0.1 and 0.5
is  that initial mean separation of particles in the initial disks of 
these models are smaller owing
to the larger number of particles.  
 The time integration of the equation of motion
 is performed by using 2-order
 leap-flog method with time step interval equal to 0.01 in 
our units. 
Energy and angular momentum  are conserved
within 1 percent accuracy in a test collisionless merger simulation.
All of the  calculations are set to be stopped at  28 $t_{\rm dyn}$.
Considering the time-scale of the present calculations and the total particle
number used in this study, it is considerably reasonable that 
the two body relaxation due to the finite particle number of
numerical simulations has a negligible effect
on  dynamical evolution of galaxy mergers in  the present study.

\placetable{tbl-1}

\subsection{Main points of analysis}
 In the present paper, we investigate the most extensively
 fundamental importance  of the rapidity of star formation ($C_{\rm SF}$)
in determining structural  and kinematical properties 
of merger remnants.
The reason for our investigation of the importance of the $C_{\rm SF}$
and the parameterization  of structural
and kinematical nonhomology of merger remnants are presented as
follows.

\subsubsection{Importance of the rapidity of star formation}

In the present study, the origin of the luminosity-dependent
structure and kinematics of elliptical galaxies
are considered to be closely associated with
the difference in star formation history between elliptical
galaxies formed by dissipative galaxy merging with star formation.
The reason for the luminosity-dependent 
structure and kinematics of elliptical galaxies can be explained 
in the present study as follows.
First, the star formation history of galaxy mergers
determines structure and kinematics of merger remnants.
Next, the star formation history of galaxy mergers depends basically
on galactic luminosity.
Structure and kinematics of elliptical galaxies  
therefore depend on galactic luminosity.
The adopted `working hypothesis' on 
the luminosity-dependent 
star formation history of elliptical galaxies (or galaxy mergers)
is considerably  realistic and
reasonable, since a growing number of observational studies support 
this hypothesis.
For example, the line ratio of [Mg/Fe], which can be interpreted as 
the strength of the past  activity of type-II supernovae relative to that of
the type I supernovae,  implies that  more luminous elliptical galaxies
are likely to truncate their star formation earlier (e.g., Worthey et al. 1992).
Furthermore, as is suggested by the analysis of $H \beta$ line index, 
the epoch, the strength, and the duration of the latest star formation
of elliptical galaxies seem quite diverse,
implying that the classical single burst picture  of elliptical galaxy formation
becomes less attractive (e.g., Faber et al. 1995).
It should be also emphasized that the $B-H$ color in more
luminous disk galaxies,
which are considered to be merger precursors of more luminous ellipticals
in the present study,
is found to be redder than the less luminous ones (Gavazzi \& Scodeggio 1996).
These lines of observational evidences  strongly motivate us to clarify
the important roles of star formation history in determining the structural 
and kinematical properties 
in elliptical galaxies formed by galaxy merging.

 In the present paper, we focus particularly on the time-scale of gas consumption
by star formation $relative$ $to$ $the$ $dynamical$ $time-scale$ $of$ 
$galaxy$ $mergers$ ($C_{\rm SF}$): 
It should be emphasized here that  
not the simple time-scale of gas consumption by star formation 
but the time-scale of gas consumption
by star formation $relative$ $to$ $the$ $dynamical$ $time$-$scale$ 
($C_{\rm SF}$) is 
considered to be 
an important determinant for structure and kinematics of merger remnants.
The details concerning  the physical meanings of the $C_{\rm SF}$ and 
the possible luminosity dependence of the $C_{\rm SF}$ are given in
the Appendix A.
We examine the most extensively
the $C_{\rm SF}$ dependence of the following three properties
of merger remnants:
Projected  density profile,
the relative importance of the global rotation in kinematics,
and  the ratio of total dynamical mass
to  luminosity mass (mass-to-light ratio).
All of these three properties are generally considered to be
fundamental factors
for  generating structural and kinematical  nonhomology of 
elliptical galaxies.
For convenient,
these three properties of merger remnants are  hereafter referred to as
``nonhomology parameters''.
Since the parameter $C_{\rm SF}$ is assumed to depend primarily
on galactic luminosity in the present study,  examining the $C_{\rm SF}$ dependence of
the above three nonhomology parameters corresponds to 
clarifying  the luminosity-dependent structure and kinematics
of merger remnants.
Thus,  clarifying  the $C_{\rm SF}$ dependence of the three
nonhomology parameters enables us to explore the origin
of the FP, in particular, a  close physical relation between
the star formation history of merger progenitor and
the origin of the FP slope.

\subsubsection{Parameterization of structural and kinematical nonhomology}

In order to investigate the dependence of the three nonhomology
parameters on the  $C_{\rm SF}$ in galaxy mergers,
we must quantify more explicitly the  nonhomology parameters.
  First of all, we  observe how  the three parameters 
    should depend on galactic luminosity in order to satisfy
    the FP slope. 
    The scaling
    relation expected from virial theorem is given
    as follows (\cite{djo88,dcd89,ps96}):
    \begin{equation}
    k_{\rm K} {\sigma_{0}}^2 \propto k_{\rm S}
    L^{1/2} {I_{\rm e}}^{1/2} (M_{\rm t}/L)
    \end{equation}
    where $M_{\rm t}$, $L$, $I_{\rm e}$, and $\sigma_{0}$ are total mass,
    luminosity,
    mean surface luminosity, and central velocity dispersion of
    elliptical galaxies.
These three terms, $k_{\rm S}$, $k_{\rm K}$,  and $M_{\rm t}/L$
are referred to as nonhomology parameters in the present study.
The quantities  $ k_{\rm S}$ and $ k_{\rm K}$  in the above equation
represent
characteristics of
mass distribution 
and those  of kinematical structure, 
    respectively.
    To be specific, 
    $ k_{\rm S}$ reflects systematical departures from universal
    density profile $R^{1/4}$ law and
    $ k_{\rm K}$ reflects the relative importance of
    global rotation in the kinematics of elliptical galaxies.
The $M_{\rm t}/L$ term basically represents both the nature of
stellar populations and the ratio of dynamical mass to luminous mass
in galaxies.
Since we here neglect the effects of stellar populations on
the  $M_{\rm t}/L$, the  $M_{\rm t}/L$ is considered to depend
predominantly   on the  ratio of dynamical  mass to luminous mass.
    By comparing the above equation (2) with  the scaling relation implied 
    by FP, we can constrain the dependence of 
    $k_{\rm S}$,  $k_{\rm K}$, and  
$M_{\rm t}/L$ on $L$, $I_{\rm e}$, and $\sigma_{0}$.
    Assuming that  $k_{\rm S}$, $k_{\rm K}$,  and $M_{\rm t}/L$ depend
    only on  $L$, we can derive the following constraint on
    these three terms.

    \begin{equation}
      K_{\rm FP} \propto  {\sigma_{0}}^2/R_{\rm e}/I_{\rm e} \propto  
       k_{\rm S} {k_{\rm K}}^{-1}   M_{\rm t}/L \propto L^{\alpha} 
    \end{equation}
where  $R_{\rm e}$ is effective radius and  
$K_{\rm FP}$ represents ``total'' nonhomology of galaxies,
which includes the above three different types of nonhomology.
If elliptical galaxies do not have non-homologous nature,
the  $K_{\rm FP}$, which corresponds to the conventionally used $M/L$,
is constant for all elliptical galaxies.
We use the  $K_{\rm FP}$  in discussing the origin of the FP
slope, as is described later. 
 $\alpha$ is required to be about 0.14  
  for $K$ band FP relation (Pahre et al. 1995; Djorgovski, Pahre, \& 
  de Carvalho 1996).
Thus the three nonhomology parameters must cooperate to
satisfy the equation (3) for the generation of  the FP.

 Next we quantify  these three terms (nonhomology
parameters) more explicitly as follows:
    \begin{equation}
     k_{\rm S} \propto R_{\rm e}/R_{\rm g} 
    \end{equation}
    \begin{equation}
      k_{\rm K} \propto  (1.0+ c_{v} \times {(v_{m}/{\sigma}_{0})}^{2}) 
    \end{equation}
    \begin{equation}
      k_{\rm M} \propto M_{t}/L \propto M_{\rm t}/M_{\rm b}  
    \end{equation}
    In the above equation (4), the $R_{\rm g}$ represents the gravitational
radius of a galaxy and is defined as $G{M_{\rm t}}^{2}/|W|$,
where $G$, $W$, and $M_{\rm t}$ are gravitational
constant, total potential energy,  and total mass in a galaxy,
respectively.
Here we adopt the relation, $k_{\rm S}  \propto R_{\rm e}/R_{\rm g}$,
rather than the relation, $k_{\rm S}  \propto
0.011 {(n-1.7)}^{2} + 0.86$, 
derived by Prugniel \& Simien (1996) 
in estimating the  $k_{\rm S}$.
The first reason for this is that the present merger model is initially
$two$ components system (dark matter and luminous component) to 
which the semi-analytic expression,  $k_{\rm S}  \propto
0.011 {(n-1.7)}^{2} + 0.86$,
for $one$ component system 
can not be so simply applied. 
The second reason is that
the $R_{\rm e}/R_{\rm g}$ is more feasible to
estimate with smaller numerical uncertainty in the present study.
The details of  the  dependence of the $R_{\rm e}/R_{\rm g}$
on the exponent $n$ in the $R^{1/n}$ light profile are 
given in Ciotti (1994). 
  In the  equation (5),
  $v_{\rm m}$ represents the maximum
  rotational velocity in a galaxy,
and the value of the parameter $c_{v}$
depends on the details of radial distribution
of kinematical properties of galaxies.
In the present study, we adopt 0.81 for the $c_{v}$,
which is the same as that of  
 Prugniel \& Simien (1996).
    The equation (5) means that $ k_{\rm K}$ is larger for elliptical
    galaxies  which are 
dynamically supported more strongly by global rotation.
    In the equation (6), $ M_{\rm b} $ represents the total luminous
    (baryonic) mass.
    We here neglect the effects of stellar populations on the
    luminosity dependence of $M_{b}/L$
    principally  because we
    discuss mainly the FP  in the context
of structural and kinematical nonhomology of elliptical galaxies. 
Presumably, the above semi-analytic expressions for the three nonhomology 
parameters are not  so  sophisticated  enough to describe
structural and kinematical properties  of elliptical galaxies
$fully$ $self$-$consistently$.
We believe, however, that
as long as we discuss global characteristics of structure and kinematics in 
elliptical galaxies  qualitatively,
the expressions are appropriate enough to describe structural
and kinematical nonhomology of elliptical galaxies.

The values of the three nonhomology parameters for merger remnants
are estimated as follows.
Since all merger models 
reaches dynamical equilibrium within  $T$ = 28.0 in our units
in the present study, we estimate the 
$ k_{\rm S}$, $ k_{\rm K}$,
and   $ k_{\rm M}$ for each of three projections ($xy$, $xz$, and $yz$ plane)
in each merger remnant at $T$ = 28.0.
Effective radius ($R_{\rm e}$) is set to be 
exactly the same as half mass radius of stellar component
in the present study. 
Projected values of $v_{\rm m}$ and $\sigma_{0}$ are measured
inside a slit with the length of 0.2 $R_{\rm e}$ for each of three projections 
in each model.
Typical length of the slit ranges from 0.04 to 0.08 in our units for the
preset study.
The value of  $v_{\rm m}$ is estimated within $R_{\rm e}$ for each of
three projections in each model.
$M_{\rm t}/M_{\rm b}$ term in the $ k_{\rm M}$ 
is estimated by dividing the total mass of the remnant
(including dark halo,
stellar,  and gaseous component) by the mass  of stellar component within
$R_{\rm e}$ for  each of three projections in each merger remnant. 
We estimate the $K_{\rm FP}$
by using the relation, $K_{\rm FP} \propto  {\sigma_{0}}^2/R_{\rm e}/I_{\rm e} $,
in the equation (3).
We use the derived vales of the  $K_{\rm FP}$ in discussing the origin of the FP slope
of elliptical galaxies formed by dissipative galaxy merging.

\placefigure{fig-1}

\section{Results}
The organization of the present investigation is two-fold.
We first numerically investigate the dependence 
of  the three nonhomology parameters, $ k_{\rm K}$,
$ k_{\rm S}$, and   $ k_{\rm M}$ on $C_{\rm SF}$ in \S 3.1.
In this \S 3.1, we furthermore observe whether  the variety of initial conditions of 
galaxy mergers such as the orbital energy,  angular  momentum,
and gas mass fraction in merger progenitor strengthens  or lessens the derived
$C_{\rm SF}$ dependence of structural and kinematical properties of merger remnants.
Second, based upon the derived  $C_{\rm SF}$ dependence 
of the above nonhomology parameters, we discuss in \S 3.2
how the rapidity of star formation ($C_{\rm SF}$)
$should$ depend on the galactic luminosity ($L$)
for the successful reproduction of the FP slope.
In this \S 3.2, we also discuss the origin of the tightness in the FP
within a frame of merger scenario for elliptical galaxy formation.

\placefigure{fig-2}

\placefigure{fig-3}

\placefigure{fig-4}

\subsection{Dependence of nonhomology parameters on the rapidity of star formation}

\subsubsection{Fundamental roles of  the $C_{\rm SF}$}
 We first observe how  the rapidity of star formation
 generates  structural and kinematical
nonhomology in merger
remnants.
 We begin to present the results of five models with different $C_{\rm SF}$
(= 7.0, 3.5, 1,75, 0.7, and 0.35) 
in  Model 1, since in these models we can observe 
 typical roles  of the 
$C_{\rm SF}$ in controlling  structural and kinematical properties of merger remnants.
Figure 1 describes the time evolution of gas mass fraction of galaxy mergers for five
models with different $C_{\rm SF}$ in Model 1.
As is  shown in this figure, the interstellar gas is more rapidly consumed by star
formation for models with larger $C_{\rm SF}$.
This typical behavior of the $C_{\rm SF}$ in
the star formation history of galaxy mergers is found to hold for all sets
of models (Model 1 $\sim$ 14) in the present
study.
 Figure 2 describes the $C_{\rm SF}$ dependence of the nonhomology parameters,
$k_{\rm S}$, $k_{\rm K}$, 
and $k_{\rm M}$ in the merger remnants at $T$ = 28.0 in our units
for  models in Model 1.
In order to show more clearly  characteristics of
the $C_{\rm SF}$ dependence of  $k_{\rm S}$, $k_{\rm K}$, and $k_{\rm M}$,
we preset   
 the best fitted line 
derived by least square fitting of each set of experimental data
to assumed relations like as  $k_{\rm S}$ ($k_{\rm K}$ and
$k_{\rm M}$)  $\propto$  
${C_{\rm SF}}^{x}$,
where the exponent $x$ is described below.
   We can observe clear trends in the $C_{\rm SF}$ dependence
   of $k_{\rm S}$, $k_{\rm K}$, and $k_{\rm M}$ as follows.
   First, as is shown in the  top panel of Figure  2, $k_{\rm S}$
   is appreciably  larger for models with larger $C_{\rm SF}$.
   This result demonstrates that the rapidity of star formation of galaxy
mergers greatly affects the mass distribution of merger remnants,
in particular, 
the projected density profile of stellar component. 
The reason for this dependence is that
the rapidity of star formation of mergers basically  determines
 how much amount of stellar component is
 transferred to the central region of the remnants during mergers,
which is a key factor for final mass distribution of stellar component in
merger remnants.
 The dependence of $k_{\rm S}$ on $C_{\rm SF}$ is described
 as $k_{\rm S}$ $\propto$ ${C_{\rm SF}}^{0.25}$.
Second,  we can observe in the middle  panel of Figure 2 
   that the larger the $C_{\rm SF}$ is, the smaller the $k_{\rm K}$ is.
   This result indicates that  galaxy mergers  with rapid gas consumption 
   become  elliptical galaxies
  which are dynamically supported  by global rotation less strongly.
   This is because total amount  of gaseous dissipation during 
   merging, which is smaller for models with larger $C_{\rm SF}$,
   principally determines how efficiently
initial total potential energy of a galaxy merger
can be converted into rotational
energy rather than random kinetic energy during galaxy merging 
and thus how strongly the merger remnant is dynamically supported
   by global rotation.
   The dependence of $k_{\rm K}$ on $C_{\rm SF}$ is described
   as $k_{\rm K}$ $\propto$ ${C_{\rm SF}}^{-0.08}$.
  Third, $k_{\rm M}$ is larger for models with larger $C_{\rm SF}$.
  This result indicates that  galaxy mergers with rapid star
  formation become ellipticals with less strongly self-gravitation
 (corresponding
to larger $k_{\rm M}$).
   This is because in the model with larger $C_{\rm SF}$,
less amount of stellar component is
   transferred to the central region
   owing to  
   less amount of kinetic energy  dissipated 
   away by gaseous dissipation during merging.
   The dependence of $k_{\rm M}$ on $C_{\rm SF}$ is described
   as $k_{\rm M}$ $\propto$ ${C_{\rm SF}}^{0.11}$.

What we should emphasize regarding the above results is that the exponent in
the $C_{\rm SF}$ dependence for
three terms, $k_{\rm S}$, ${k_{\rm K}}^{-1}$, and $k_{\rm M}$ 
are all positive.
Specifically, merger remnants with larger ratio of effective radius to
gravitational radius are more likely to show both the smaller degree of 
global rotation and the smaller degree of self-gravitation of stellar component.
Since 
$K_{\rm FP} \propto
k_{\rm S} 
{k_{\rm K}}^{-1} k_{\rm M}$,
this result means that
these three terms cooperate to make the 
$C_{\rm SF}$ dependence of  $K_{\rm FP}$  in  merger remnants become stronger.
This cooperation of three nonhomology parameters can be more clearly 
observed in  Figure 3 in which the $k_{\rm S}$ dependence of
${k_{\rm K}}^{-1}$ and $k_{\rm M}$ is described.
As is shown in this figure, $both$ the ${k_{\rm K}}^{-1}$ and $k_{\rm M}$ 
are larger for models with larger  $k_{\rm S}$.
This remarkable cooperation observed in the $C_{\rm SF}$ dependence
of nonhomology parameters, $k_{\rm S}$, $k_{\rm K}$, and $k_{\rm M}$,
is first demonstrated in the present dissipative merger model, which is 
accordingly
one of successes in the present study.
As a natural result of the cooperation,
we can clearly observe 
the $C_{\rm SF}$ dependence of  the $K_{\rm FP}$ for models in the Model 1
in  Figure 4.
The dependence of $K_{\rm FP}$ on $C_{\rm SF}$ is described
as $K_{\rm FP}$ $\propto$ ${C_{\rm SF}}^{0.34}$.
Thus the origin of structural and kinematical nonhomology of elliptical galaxies 
is demonstrated to be closely associated with the difference 
in the rapidity of star formation of galaxy mergers.

\placefigure{fig-5}

\placefigure{fig-6}

\placefigure{fig-7}

\placefigure{fig-8}

\placefigure{fig-9}

\subsubsection{Importance of  other parameters}

 Next we observe whether diversity in initial conditions of galaxy merging
 strengthens or lessens the 
$C_{\rm SF}$ dependence of $k_{\rm S}$, $k_{\rm K}$, and $k_{\rm M}$ 
derived in \S 3.1.1.
In particular, we investigate how the diversity of initial intrinsic spins,
orbital energy and angular momentum, gas mass fraction at the epoch of
galaxy merging, ratio of dynamical mass to luminous mass 
introduces a scatter in the $C_{\rm SF}$ dependence of $k_{\rm S}$, $k_{\rm K}$,
and $k_{\rm M}$.

In order to show more clearly the results,
we divide the 14 sets of models (Model 1 $\sim$ 14)
into five  groups of models (Group A 
$\sim$ E).
Group A includes four sets of models, Model 1, 2, 3, and 4,
in which initial intrinsic spins of two disks ($\theta_{i}$ and $\phi_{i}$)
are different from models
to models.
In this group, we investigate  
how the difference of
$\theta_{i}$ and $\phi_{i}$ 
 can affect the $C_{\rm SF}$
dependence of $k_{\rm S}$, $k_{\rm K}$, and $k_{\rm M}$. 
Group B includes six sets of models, Model 1, 5, 6, 7, 8, and 9
in which initial orbital energy and angular momentum
($e_{\rm p}$ and $r_{\rm p}$) are different from models
to models.
In this group, we investigate  
how the difference of
$e_{\rm p}$ and $r_{\rm p}$ 
can affect the $C_{\rm SF}$
dependence of $k_{\rm S}$, $k_{\rm K}$, and $k_{\rm M}$. 
Group C  includes three  sets of models, Model 1, 10, and 11,
in which maximum speed of rotation curve of
initial disks  ($V_{\rm max}$) is  different from models
to models.
In this group, we investigate  
how the difference of
$V_{\rm max}$ 
can affect the $C_{\rm SF}$
dependence of $k_{\rm S}$, $k_{\rm K}$, and $k_{\rm M}$. 
Group D  includes four sets of models, Model 1,  12, and 13,
in which gas mass fraction ($M_{\rm g}$)  are different from models
to model,  and Model 14 in which the epoch of galaxy merging
is delayed (models with larger $r_{\rm in}$).
In this group, we investigate  
how the difference of gas mass fraction at the epoch
of galaxy galaxy merging can affect the $C_{\rm SF}$
dependence of $k_{\rm S}$, $k_{\rm k}$, and $k_{\rm M}$. 
Group E includes all sets of models, Model 1 $\sim$ 14.
For each model group (Group A $\sim$ E),
we obtain the exponent $\alpha_{i}$ ($i$ = S, K, M, and FP) 
in the $C_{\rm SF}$  dependence of  
$k_{\rm S}$, $k_{\rm K}$, $k_{\rm M}$, and
$K_{\rm FP}$  (e.g., $k_{\rm i} 
\propto {C_{\rm SF}}^{\alpha_{\rm i}}$, $i$ = S, K, and M). 
The exponents $\alpha_{i}$ ($i$ = S, K, M, and FP)  in the $C_{\rm SF}$ dependence  
are summarized in  Table 2.
The  exponent $\alpha_{FP}$ describes the strength of the correlation between
the $C_{\rm SF}$ and non-homologous  nature of merger remnants. 
The physical meaning for $\alpha_{\rm SF}$ in the seventh
column of the Table 2 is described later.  
The $C_{\rm SF}$ dependences  
of $k_{\rm S}$, $k_{\rm K}$, $k_{\rm M}$, and $K_{\rm FP}$ for Group
A, B, C, D, and E
are given in Figures 5, 6, 7, 8, and 9, respectively.
Characteristics of the  $C_{\rm SF}$ dependence of
$k_{\rm S}$, $k_{\rm K}$, $k_{\rm M}$, and $K_{\rm FP}$
for each model group (Group A $\sim$ E) are summarized  as follows.

In the Group A, 
diversity in initial intrinsic spins of two disks 
is found  to introduce an appreciable scatter in
the values of the three nonhomology parameters for a given $C_{\rm SF}$
(See Figure 5.).
However,  characteristics of
the $C_{\rm SF}$ dependence of $k_{\rm S}$, $k_{\rm K}$, and $k_{\rm M}$
are similar to that derived for  Model 1.
The exponent $\alpha_{FP}$ in the $C_{\rm SF}$ dependence of
$K_{\rm FP}$ is positive, which is consistent with the result
derived in  Model 1.
These results indicate that 
fundamental roles of the rapidity of star formation in generating
structural and kinematical nonhomology of merger remnants do not depend
on initial intrinsic spins of progenitor disks.
These results furthermore suggest that the observed small scatter
in the scaling relation of the FP reflects 
the diversity in  initial intrinsic spins of galaxy mergers.
The value of $\alpha_{FP}$ in the $C_{\rm SF}$ dependence of
$K_{\rm FP}$ derived for this Group A is 0.41,
which is appreciably larger than that derived for Model 1.

In the Group B,
characteristics of 
the derived $C_{\rm SF}$ dependence of $k_{\rm S}$, $k_{\rm K}$, $k_{\rm M}$,
and  $K_{\rm FP}$ are similar to those derived in the Group A:
Although the diversity   
in initial orbital energy and angular momentum  
actually introduces  an appreciable scatter in
the $C_{\rm SF}$ dependences of $k_{\rm S}$, $k_{\rm K}$,  $k_{\rm M}$, 
and $K_{\rm FP}$, the $C_{\rm SF}$ dependences
are  qualitatively the same
as those derived for Model 1 (See Figure 6.). 
These results also indicate that 
the fundamental roles of the rapidity of star formation in generating
structural and kinematical nonhomology of merger remnant do not depend
on initial orbital energy and angular momentum of galaxy mergers. 
The value of $\alpha_{FP}$ in the $C_{\rm SF}$ dependence of
$K_{\rm FP}$ derived for this Group B is 0.37,
which is slightly larger than that derived for Model 1 but smaller
than that for Group A.

In the Group C, 
the difference in initial ratio of dynamical mass to luminous mass 
is found to introduce  a  larger scatter in the nonhomology parameter
$k_{\rm M}$ (See Figure 7.).
The reason for this is that the degree of self-gravitation of stellar component 
in merger remnants
is  determined by the initial degree of self-gravitation of progenitor disks
as well as by the rapidity of star formation.
This result suggests that the diversity in the degree of self-gravitation
actually observed in galactic disks (e.g. Saglia 1996)
introduces a scatter in the scaling relation of the FP
and thus that degree of self-gravitation of galactic disks
should not be so different between merger progenitor for
the considerably smaller dispersion in the FP.
The scatter in the nonhomology parameter
$k_{\rm S}$ and $k_{\rm K}$, on the other hand,
is not so larger compared with  that in the $k_{\rm M}$.
The value of $\alpha_{FP}$ in the $C_{\rm SF}$ dependence of
$K_{\rm FP}$ derived for this Group C is 0.42.

In the Group D, 
the difference in gas mass fraction of disks at the epoch of galaxy merging
($M_{\rm g}$ and $r_{\rm in}$) 
is found to  yield a  relatively
larger scatter in the three nonhomology parameters,
in particular, in the $k_{\rm K}$ (See  Figure 8.).
This is principally because in the models with smaller gas mass fraction
($M_{\rm g}$ = 0.1),
the rapidity of star formation can not affects so greatly the final
structural and kinematical  properties of merger remnants.
As a result of this,
the exponents $\alpha_{\rm S}$,  $\alpha_{\rm K}$,  and $\alpha_{\rm M}$
are all smaller compared with
those derived in other model groups, Group A,  B and C
(See the Table 2.). 
These results imply that in galaxy mergers with smaller gas mass at lower
redshift universe, the rapidity of star formation does not
have remarkable effects on structure and kinematics of merger remnants.
These results furthermore imply that 
the lower
redshift galaxy mergers introduce a larger scatter
in the scaling relation of the FP.
The value of $\alpha_{FP}$ in the $C_{\rm SF}$ dependence of
$K_{\rm FP}$ derived for this Group D is 0.22, which shows the
weakest correlation among  the five model groups (Group  A $\sim$  E).

In the Group E, 
characteristics of 
the derived $C_{\rm SF}$ dependence of $k_{\rm S}$, $k_{\rm K}$, $k_{\rm M}$,
and  $K_{\rm FP}$ are similar to those derived in Model 1:
Although the diversity   
in initial conditions of galaxy merging 
actually introduces  an appreciable scatter in
the $C_{\rm SF}$ dependences of $k_{\rm S}$, $k_{\rm K}$,  $k_{\rm M}$, 
and $K_{\rm FP}$,
the $C_{\rm SF}$ dependences are  qualitatively the same
as that derived for Model 1 (See Figure 9.). 
The value of $\alpha_{FP}$ in the $C_{\rm SF}$ dependence of
$K_{\rm FP}$ derived for this Group E is 0.36.
Since this Group E includes all dissipative merger models,
this result clearly demonstrates that the rapidity of  star formation
is a $primarily$ important parameter which governs structural
and kinematical nonhomology of merger remnants.
This result furthermore suggests that the origin of the FP slope
can be closely associated with the rapidity of star formation
of galaxy mergers.

Thus we found that although the diversity in the initial conditions
of merger progenitor can introduce an appreciable  scatter in the
$C_{\rm SF}$ dependence of nonhomology parameters,
the fundamental roles of  $C_{\rm SF}$ in generating structural
and kinematical nonhomology remain essentially the same as is 
already obtained in \S 3.1.1.
These results clearly indicate that total amount of gaseous
dissipation, which is principally controlled by the rapidity
of star formation, is an important parameter for 
structural and kinematical properties of merger remnants.
These results furthermore suggest that if  a specific
relation between the rapidity of star formation ($C_{\rm SF}$)
and galactic luminosity ($L$)  actually exists,
the origin of the FP can be  
at least qualitatively explained by the present merger model,
as is described below.

\placetable{tbl-2}

\placefigure{fig-10}

\subsection{The FP of merger remnants}
 Based upon the $C_{\rm SF}$ dependence of the $K_{\rm FP}$ derived
in \S 3.1, we here discuss how the $C_{\rm SF}$ should depend on
the galactic luminosity ($L$)  for successful reproduction of the FP
slope.
We then create the FP relation of elliptical galaxies formed by
dissipative galaxy merging by using the required  $L$ dependence of
the  $C_{\rm SF}$ and 
by assuming a specific relation between
 galactic mass (or luminosity) and typical size of galaxies.

\subsubsection{Required luminosity-dependence of the  $C_{\rm SF}$}
 
 Dependence of mass-to-light ratio on the galactic luminosity implied by
the scaling relation of the FP requires that $K_{\rm FP}$ ($\propto M/L$)
should depend on
$L$ as $K_{\rm FP} \propto L^{0.14}$ (e.g., Pahre et al. 1995; Djorgovski et al. 1996).
In the present study, for example,
the $K_{\rm FP}$ is found to depend on the $C_{\rm SF}$
as $K_{\rm FP} \propto {C_{\rm SF}}^{0.36}$ for model group E in which
all dissipative merger models are included.
By using the above two dependences of the $K_{\rm FP}$,
we can obtain the result that the $C_{\rm SF}$ should depend on 
$L$ as $C_{\rm SF} \propto L^{0.39}$ for explaining the origin of
the FP slope.
The exponent in the expected dependence of
$C_{\rm SF}$ on  $L$ ($C_{\rm SF} 
\propto L^{\alpha_{\rm SF}}$)
for other model groups
(Group A, B, C, and D) are summarized in the seventh  column of the Table 2. 
As is shown in the Table 2, the value of the expected exponent
ranges from 0.33 to 0.64.
Probably the most plausible value of the exponent in the present study
is 0.39 which is derived for model group E including all the dissipative 
merger models.
This  expected dependence of $C_{\rm SF} \propto L^{0.39}$
implies that if more luminous elliptical galaxies are formed 
by galaxy mergers with more rapid star formation,
the slope of the FP can be at least qualitatively explained
in the present merger model.
This result furthermore demonstrates that although  a specific relation
between star formation history of galaxy mergers and  galactic luminosity
is required, the origin of the FP slope can be closely associated with
the star formation history of elliptical galaxies formed by dissipative galaxy
merging.
It should be noted here that since the present merger model
is rather idealized and less realistic for $real$ dissipative galaxy
mergers,
there still remains uncertainty in the derived dependence of 
$C_{\rm SF} \propto L^{0.39}$.
What we can say in the preliminary stage  of the present study
is that although it is not clear whether or not
the derived $C_{\rm SF} \propto L^{0.39}$
is consistent totally with the observational results on the possible 
$L$ dependence of the $C_{\rm SF}$, the present numerical results strongly
suggests the importance of the $L$ dependence of  the $C_{\rm SF}$
in creating the FP of elliptical galaxies.

\subsubsection{Creation of the FP}

 By using the above relation, $C_{\rm SF} \propto L^{0.39}$,  and by assuming
a specific relation of $L \propto {R_{d}}^{2}$ (Freeman's law)
and constant ratio of dynamical mass to luminous mass in merger progenitor,
we can convert the present $nondimentional$ results into
$dimentional$ ones.
Specifically, the dimensionless effective radius,
$R_{\rm e}$, central velocity dispersion, $\sigma_{0}$,
and the surface luminous density, $I_{\rm e}$, derived 
for each of three projections
in each merger model are converted into three new dimensional FP variables,
$R_{\rm e}'= L^{0.5} R_{\rm e} =
{C_{\rm SF}}^{1.28} R_{\rm e} $, $\sigma_{0}'=
L^{0.25} \sigma_{0} = {C_{\rm SF}}^{0.64} \sigma_{0} $,
and $I_{\rm e}'=I_{\rm e} $, respectively.
By using these  FP variables derived for each model and by
assuming  ${\mu}_{\rm e}'$ = $-2.5 \log I_{\rm e}'$, 
we can create  the FP of elliptical galaxies formed by dissipative 
galaxy merging.
We adopt the same two variables ($\log R_{\rm e}'$ and
$\log \sigma_{0}' + 0.24 {\mu}_{\rm e}'$) as are used in the
$K$ band FP created by Pahre et al. (1995) and Djorgovski et al. (1996)
and plot each FP
variable in Figure 10.
In this figure, we redefine the three  dimensional variables
($R_{\rm e}'$, $\sigma_{0}'$, and  ${\mu}_{\rm e}'$)
as $R_{\rm e}$, $\sigma_{0}$, and
${\mu}_{\rm e}$ only for simplicity.  
In this figure, the results for
the models in  Model 10 and 11 in which the initial
ratio of dynamical mass to luminous mass is different from other models
are not plotted, principally because
these models in Model 10 and 11 do  not match with the adopted
assumption of constant ratio of dynamical mass to luminous mass in progenitor
disks (Actually, the results in Model 10 and 11 are  found to introduce
an  appreciably larger scatter in the reproduced FP.).
The  solid line in this figure indicates the line with the same slope as
is observed in the $K$ band FP created by Pahre et al. (1995)
and Djorgovski et al. (1996)
and with an arbitrary zero point for the present model.
The reproduced FP slope of merger remnants reflects the difference in
total amount of gaseous dissipation, which is controlled greatly
by the rapidity of star formation, among galaxy mergers. 
As is shown in this figure, 
there is an appreciable scatter in the FP.
This result probably reflects either the uncertainty of the adopted scaling
relations between the rapidity of star formation and galactic luminosity,
or  some defects of the present merger model,
such as dissipative dynamics of interstellar gas, star formation model,
and initial mass distribution and kinematics of merger progenitor.
What we should stress  here is that the reproduced FP in Figure 10
is not exactly
the same as the observed FP, the Figure
10  suggests, however, that 
if a specific physical relation between the rapidity of
star formation of galaxies 
and galactic luminosity $actually$ exists,
the FP can be reproduced in remnants of dissipative galaxy mergers with 
star formation.

 Thus, our numerical results indicate that the slope of the FP
reflects the difference in dissipative dynamics, which is probably  controlled
the most effectively by star formation history of galaxies,
between elliptical galaxies formed by galaxy merging.
Although the  importance of gaseous dissipation in reproducing the FP has
been already suggested by several authors (e.g., Djorgovski
et al.  1988;
Bender et al. 1992, 1993), the present numerical study,  however,
first demonstrated this importance in the context of star formation history of
elliptical galaxies.
The essentially important physics related to the origin of the FP
is violent relaxation combined with gaseous dissipation in the present study.
Since the star formation history of galaxies, in particular,
the rapidity of star formation, is a key determinant for the effectiveness
of gaseous dissipation, the origin of the FP can be closely associated
with the star formation history of galaxies.
It is however not clarified in the present study
whether this importance of galactic star formation history in the reproduction
of the FP can be $actually$ applied to more realistic situation of galaxy
merging which is allowed by a specific cosmology and environment. 
Accordingly we will investigate this importance more throughly by
using  more elaborated numerical simulations with more cosmologically
plausible initial orbit configuration and more realistic initial physical
conditions of merger progenitor.


\section{Discussion}
\subsection{Rapidity of star formation in  galaxy mergers and the origin of
the FP slope}
 In the previous sections, we demonstrated that the rapidity of star formation
($C_{\rm SF}$)
can greatly affect both  mass distribution and kinematics of merger remnants.
We furthermore suggested  that if a specific  dependence of 
the $C_{\rm SF}$ on  galactic luminosity ($L$) is satisfied 
like as $C_{\rm SF} \propto L^{0.39}$
among merger progenitor,
the origin of the FP can be explained at least  qualitatively
in the present merger model.
What we should stress here is that not the simple time-scale
of gas consumption  by star formation but
the ratio of the dynamical time-scale of galaxies to the   time-scale of
gas consumption  
is found to be a key determinant for the reproduction of the FP.
The reason for this importance of the time-scale ratio ($C_{\rm SF}$) 
is that the  $C_{\rm SF}$ can control the total amount of gaseous dissipation
during galaxy merging,
which is a primarily important factor for structure and kinematics of merger remnants.
This importance of the rapidity of star formation in
explaining the origin of  
the FP is first pointed out in the present study.
It has not been however clarified
whether or not the expected relation of $C_{\rm SF} \propto L^{0.39}$
is astrophysically plausible and realistic.
Accordingly, it is our next step to assess the validity of the expected $L$
dependence of
the  $C_{\rm SF}$ in order that we confirm a close physical
relation between
the rapidity of
the star formation and the origin of the FP.

To confirm the above important roles of the rapidity of star formation
in the reproduction of the FP, we must foremost know the 
dependence of  time-scale  of  gas consumption
by star formation  on  galactic luminosity.
However, 
the dependence of star formation history of galaxies
on galactic luminosity has not been so extensively examined in the previous observational
studies.
The reason for the smaller number of observational studies on the luminosity-dependent
star formation history of galaxies is  
probably  that  characteristics of star formation history in galaxies
have  been  generally
considered to be more closely associated with the Hubble type rather than the
galactic luminosity or mass (e.g., Sandage 1984; Kennicutt 1992; Larson 1992; 
Roberts \& Haynes 1994).
Only a few  observational studies regarding  the dependence of 
global color, surface density, and gas mass fraction on  galactic luminosity
in disk galaxies have been accumulated  which  suggest strongly that
galactic luminosity or mass is a primarily important parameter which can
control the whole history of galactic star formation (see e.g.,
Gavazzi \& Scodeggio 1997; McGaugh \& Blok 1997).
We accordingly await
more extensive observational studies in order to 
explore more quantitatively
the important roles of the rapidity of star formation in the creation of the 
FP.
Thus, it is still not clear whether the rapidity of star formation
$actually$ plays  a vital role in creating the FP,
the present numerical study, however,   suggests strongly
that both more extensive
theoretical and observational studies concerning 
the dependence of galactic star formation history on
galactic luminosity are doubtlessly worthwhile for
better understanding of the origin of the FP.

\subsection{Comparison  with previous works}
 There are a number of previous studies in which main points on
the origin of the FP slope are greatly different from those of
the present study.
Hjorth \& Madsen (1995) 
constructed a self-consistent equilibrium model
of elliptical galaxies 
and demonstrated that if the more luminous 
elliptical galaxies have the so-called  $R^{1/n}$ light profile with larger $n$, the
slope of the FP can be successfully reproduced. 
They furthermore discussed that the origin of the FP slope can be  understood
in terms of dissipationless galactic dynamics, in particular,  
the physics of violent relaxation.
Capelato et al. (1995) 
found that initial orbital energy of merging systems is 
a primarily important factor for the radial profile of
velocity dispersion and mass distribution of merger remnants,
and thus for successful reproduction of the FP slope. 
Based upon this result, they suggested that if the more luminous
elliptical galaxies are formed by galaxy merging with larger initial
orbital energy, the FP can be reproduced successfully.
Levine (1997) 
discussed how the three basic variables in
the FP should change in dissipationless merger events
for the reconstruction or  maintenance of the FP 
of elliptical galaxies formed by  galaxy merging. 
The essential differences between the above previous studies
and the present one are the following two. 
First, the dissipationless dynamics of galaxy merging is
a primarily important factor in the previous works whereas
the dissipative dynamics combined with  galactic star formation 
is essential for the present merger model.
Second, the origin of the FP slope can be essentially ascribed
to the difference of initial conditions of galaxy mergers with
different luminosity or mass in the previous studies 
whereas the difference in
star formation history between  merging galaxies is reflected on
the FP slope in the present merger model.
Since both  the above previous studies and the present one
are still
in the preliminary stage of exploring the origin of the FP, 
it is safe for us to say that both merger models are actually promising
candidates which provide valuable clues about  the origin of the FP. 
What the above dissipationless and the present dissipative merger models
should foremost address in the future
studies is on whether these models can $self$-$consistently$
explain
$other$ $fundamental$ $characteristics$ of elliptical galaxies such as
the color-magnitude relation of elliptical galaxies. 
Both the dissipationless and dissipative models  at least in
the preset stage seem to
have a number of disadvantages
or caveats for explaining both the origin of the FP and that of
the other universal scaling relations of elliptical galaxies.
Further theoretical studies based on numerical simulation with variously
different and astrophysically plausible initial conditions will
point out more clearly advantages and disadvantages
in  explaining the origin of the FP
for each of the two models.

\subsection{Future work}

 Our final goal is to construct a self-consistent
model of elliptical galaxy formation
which can explain not only  the origin of the FP but also other fundamental 
characteristics  observed in elliptical galaxies, such as the color-magnitude (CM)
relation (Faber 1973; Visvanathan \& Sandage 1977),
 $\rm Mg_{2} - \sigma$ relation (Burstein et al. 1988), age-metal-conspiracy
 in stellar populations (Faber et al. 1995; Worthey et al. 1996), Kormendy-relation
(e.g., Djorgovski et al. 1996), and
dichotomy between boxy and disky elliptical galaxies (Kormendy \&
Bender 1996).
Thus it is essential for our future studies to confirm whether
the derived relation between the  rapidity of star formation and the 
galactic luminosity for the reproduction of the FP slope ($C_{SF} \propto 
L^{0.39}$)
can also be essentially important  for  understanding
  the origin of the above fundamental characteristics of elliptical
galaxies.
In particular,  it is  the most important  future study to
discuss  the validity of the derived $L$ dependence of $C_{SF}$ in
explaining also the origin of the CM relation,
which is  one of the most fundamental universal relations of elliptical galaxies. 
Actually we have already obtained 
the result that even the origin of the CM 
relation can be closely associated with the rapidity of star formation of
galaxy mergers (Bekki \& Shioya 1997).
However it has not yet been clarified throughly whether the $L$ dependence
of the $C_{SF}$ required for the reproduction of the FP is admittedly
similar to that required for the CM relation.
Accordingly, we will investigate how the rapidity of 
star formation ($C_{SF}$) should depends on the galactic luminosity ($L$)
in order that both the origin of the FP and that of the CM relation
can be reproduced successfully for the present merger model
in a self-consistent  manner. 
Although the rapidity of star formation is probably not only the parameter which
can explain the origin of fundamental scaling relations of elliptical galaxies,
we consider that it is doubtlessly worthwhile
to investigate  the importance of the rapidity of star formation
in generating scaling relations other than
the FP in elliptical galaxies.

\section{Conclusion}
 We have numerically investigated  structural
and kinematical properties of  remnants of galaxy mergers between
late-type  spirals considerably abundant in interstellar gas in order  
to explore the origin of the fundamental plane (FP)
of elliptical galaxies.
We particularly considered  that the origin of the slope of the FP  is essentially
due to the nonhomology in structure and kinematics of  elliptical galaxies
formed by dissipative galaxy merging with star formation.
We found that the rapidity of star formation, which is defined as the ratio
of dynamical time-scale of merger progenitor 
to the time-scale of gas consumption by  star formation, 
is a key determinant for nonhomology parameters
such as the density profile, the degree of global rotation,
and the ratio of total dynamical mass to luminous mass,
in merger remnants.
This result strongly suggests that  structural and kinematical
nonhomology in elliptical galaxies
can be  closely associated with the difference in  star formation history 
between elliptical galaxies formed by  dissipative galaxy merging.
This result implies  that if a specific relation between the rapidity of
star formation ($C_{\rm SF}$) and galactic luminosity ($L$) is satisfied
like as $C_{\rm SF} \propto L^{0.39}$, the required
dependence of mass-to-light ratio on galactic mass in the
FP  (the slope of the FP)
can be explained  at least qualitatively in the present
merger model.
Further extensive observational studies  probably  assess the validity of the 
expected $L$ dependence of the $C_{\rm SF}$ and thus determine 
whether or not the star formation history of merger progenitor galaxies
can be a crucial factor for the creation of the FP. 
Although the present merger model is indeed rather idealized and less realistic
for $real$  dissipative galaxy mergers with star formation  at higher redshift universe, 
it appears to have succeeded in demonstrating  that the origin of the FP
can reflect a close physical relation between galactic luminosity 
and star formation history of
merger progenitor galaxies, in particular, the rapidity of star formation
of galaxies.
What we should emphasize here is that the proposed idea is  only
one of several 
promising ones in which physical meanings of the FP are extensively
examined in variously different ways.
Probably, both the slope and the tightness of the FP contain more profound
meanings on the formation and evolution of elliptical galaxies than
we considered in the present study.
Thus, it is  our future study to  explore another  promising
candidates related to  the  origin of the FP,
such as the  age and metallicity difference of elliptical
galaxies and  hierarchy of galaxy merging.

\acknowledgments

We are grateful to the referee for valuable comments, which contribute to
improve the present paper.
K.B. thanks to the Japan Society for Promotion of Science (JSPS) 
Research Fellowships for Young Scientist.

\appendix
\section{The expected luminosity dependence of $C_{\rm SF}$}

The reason for the importance of the $C_{\rm SF}$, which represents
the time-scale of gas consumption relative to dynamical time-scale,
is explained as follows.
Redistribution of mass and angular momentum
of galaxy mergers, which is a crucial factor for structure and kinematics
of merger remnants, 
can proceeds $within$ $only$ $an$ $order$ $of$ $dynamical$ $time$
 in galaxy mergers. 
Dissipative dynamics of interstellar gas during galaxy merging
greatly affects the redistribution of mass and angular momentum
in galaxy mergers, as is demonstrated by extensive numerical
simulations of
Barnes \& Hernquist (1992, 1996).
Accordingly it is  essential how greatly the gaseous dissipation
can affect the redistribution process $within$ $an$ $order$ $of$ $the$ $dynamical$
$time-scale$ $of$ $galaxy$ $mergers$.
Since, the time-scale of gas consumption by star formation
basically determines the time-evolution of gas mass which is a key
determinant for dissipative dynamics of interstellar gas,
the ratio of 
the gas consumption time-scale
to the dynamical time-scale 
is thus more essential for structure and kinematics of merger remnants.
For example, if the dynamical time-scale is much larger than 
the gas consumption  time-scale,  star formation can proceed quite efficiently
before the system reach the dynamical equilibrium,  and consequently 
gaseous dissipation does not play a great role in determining
the structural and kinematical properties of merger remnants.
Thus, since the gaseous dissipation
can strongly affect the dynamical 
evolution of mergers which proceeds within the order of dynamical time-scale, 
the ratio of the above two time-scales
are expected to be more essential for the structural and kinematical properties
of merger remnants.
We should emphasize here
that  for convenient, the $inverse$ ratio of the time-scale of gas consumption
by star formation to the dynamical time-scale is referred to as the rapidity
of star formation.
We consider the $C_{\rm SF}$ to depend basically on galactic luminosity,
as is described below.

It has not been observationally clarified how the $C_{\rm SF}$ actually
depends  on galactic luminosity.
Accordingly we here give a simple order estimation
for the luminosity dependence of the $C_{\rm SF}$.
The expected luminosity (mass) dependence of  the $C_{\rm SF}$ is described 
as follows.
The parameter $C_{\rm SF}$ is set to be proportional to 
 $T_{\rm dyn}$/$T_{\rm SF}$,
where  $T_{\rm dyn}$ and $T_{\rm SF}$  are the  dynamical time-scale
and the time-scale of gas consumption by star formation,
respectively.
We here define  the mass of a galactic disk, the total mass   of 
luminous  and dark matter,  and size of the progenitor as 
$M_d$, $M_t$, and $R_d$, respectively.
We consider here 
that gas mass in a disk is equal to $M_d$ for simplicity. 
The $T_{\rm dyn}$ is given as  
\begin{equation}
T_{\rm dyn} \propto R_d^{3/2} M_t^{-1/2}
\end{equation}
Provided that the coefficient in the Schmidt law is not dependent on the
galactic mass (or luminosity), we can derive $T_{\rm SF}$ as follows. 
\begin{equation}
T_{\rm SF} \propto \Sigma^{1 - \gamma} \; ,
\end{equation}
where $\Sigma$ is the surface density of the gas disk. 
The parameter $\gamma$ is the exponent of Schmidt law, which is the same as 
that used in previous subsections. 
Assuming the Freeman's law and the constant ratio of $R_d$ to 
the scale length of exponential disk, 
we derive
\begin{equation}
\Sigma \propto M_d R_d^{-2} \sim {\rm const.} 
\end{equation}
Assuming that the degree of self-gravity of a galactic disk is described as 
\begin{equation}
M_t \propto M_d^{(1 - \beta)} \; \; ,
\end{equation} 
then we can derive 
\begin{equation}
C_{\rm SF} \propto M_d^{1/4 + \beta/2} \; \; .
\end{equation}
Since $\beta$ is considered to have positive value 
($\beta = 0.6$: Saglia 1996), 
this relationship predicts that $C_{\rm SF}$ becomes larger as 
$M_d$ increases.
Furthermore, if we  
assume the constant mass to light ratio for luminous matter,
we can  obtain the luminosity dependence of $C_{\rm SF}$. 
Alternatively,
if we adopt the observed trend that more luminous disks have large surface 
density, such as $\Sigma \propto M_d$ (McGaugh \& Blok 1997), we can obtain
$C_{\rm SF} \propto M_d^{1/2 + \beta/2}$.
These simple theoretical arguments   accordingly suggest that larger (or more luminous)
disk galaxies are more likely to have larger values of $C_{\rm SF}$.
It should be noted here that the derived $L$ (or $M_{\rm d}$) dependence of
the $C_{\rm SF}$ is not necessarily plausible (or realistic)
for $real$ galaxies owing to the lack of extensive observational studies
on the luminosity dependence of the $C_{\rm SF}$.

\clearpage

\begin{deluxetable}{ccccccccccc}
\footnotesize
\tablecaption{Model parameters 
\label{tbl-1}}
\tablewidth{0pt}
\tablehead{
\colhead{Model no.}  & \colhead{$V_{\rm max}$} & 
\colhead{$M_{\rm t}$} & \colhead{$M_{\rm g}$} & 
\colhead{$r_{\rm in}$} & \colhead{$r_{\rm p}$} &
\colhead{$e_{\rm p}$} & \colhead{$\theta_{1}$} &
\colhead{$\theta_{2}$} & \colhead{$\phi_{1}$} &
\colhead{$\phi_{2}$}  }  
\startdata

    1 & 1.81 & 3.80  & 1.00  & 6.00  & 1.00  & 1.00 & 30.0 & 120.0 &  
90.0  & 0.0   \\ 

   2 & 1.81 & 3.80  & 1.00  & 6.00  & 1.00  & 1.00 & 0.0 & 120.0 &  
0.0  & 0.0   \\ 

   3 & 1.81 & 3.80  & 1.00  & 6.00  & 1.00  & 1.00 & 0.0 & 60.0 &  
0.0  & 0.0   \\ 

   4 & 1.81 & 3.80  & 1.00  & 6.00  & 1.00  & 1.00 & 180.0 & 120.0 &  
0.0  & 0.0   \\ 

   5  & 1.81 & 3.80  & 1.00  & 6.00  & 0.50  & 1.00 & 30.0 & 120.0 &  
90.0  & 0.0   \\ 

   6 & 1.81 & 3.80  & 1.00  & 6.00  & 1.25  & 1.00 & 30.0 & 120.0 &  
90.0  & 0.0   \\

  7  & 1.81 & 3.80  & 1.00  & 2.00  & 1.00  & 0.00 & 30.0 & 120.0 &  
90.0  & 0.0   \\ 

 8  & 1.81 & 3.80  & 1.00  & 6.00  & 1.13  & 0.78 & 30.0 & 120.0 &  
90.0  & 0.0   \\ 

 9   & 1.81 & 3.80  & 1.00  & 6.00  & 0.92  & 1.18 & 30.0 & 120.0 &  
90.0  & 0.0   \\ 

  10   & 1.28 &  1.99  & 1.00  & 6.00  & 1.00  & 1.00 & 30.0 & 120.0 &  
90.0  & 0.0   \\ 

   11 & 2.56 & 7.42  & 1.00  & 6.00  & 1.00  & 1.00 & 30.0 & 120.0 &  
90.0  & 0.0   \\ 

   12  & 1.81 & 3.80  & 0.50  & 6.00  & 1.00  & 1.00 & 30.0 & 120.0 &  
90.0  & 0.0   \\ 

   13  & 1.81 & 3.80  & 0.10  & 6.00  & 1.00  & 1.00 & 30.0 & 120.0 &  
90.0  & 0.0   \\ 

   14  & 1.81 & 3.80  & 1.00  & 18.00  & 1.00  & 1.00 & 30.0 & 120.0 &  
90.0  & 0.0   \\ 

\enddata

\end{deluxetable}

\begin{deluxetable}{ccccccc}
\footnotesize
\tablecaption{Summary of the $C_{\rm SF}$ dependence of nonhomology
parameters.
\label{tbl-2}}
\tablewidth{0pt}
\tablehead{
\colhead{Group}  & \colhead{Model number} & 
\colhead{$\alpha_{\rm S}$} & 
\colhead{$\alpha_{\rm K}$} & 
\colhead{$\alpha_{\rm M}$} & 
\colhead{$\alpha_{\rm FP}$} & 
\colhead{$\alpha_{\rm SF}$} } 
\startdata

A & 1,2,3,4  & 0.11  & -0.09  & 0.25 & 0.41 & 0.34 \\ 
B & 1,5,6,7,8,9  & 0.11  & -0.11  & 0.25  & 0.37 & 0.38 \\ 
C & 1,10,11  &  0.12 & -0.14  & 0.28 & 0.42 & 0.33 \\ 
D & 1,12,13,14  & 0.06  & -0.06  & 0.13  & 0.22 & 0.64  \\ 
E & all models   & 0.10   & -0.10  & 0.23 & 0.36 & 0.39  \\ 

\enddata

\end{deluxetable}

\clearpage

\figcaption{Time evolution of gas mass   fraction in  
galaxy mergers for models 
with $C_{\rm SF}$ = 7.0, 3,5, 1.75, 0.7, and 0.35
in Model 1.
Note that interstellar gas is more rapidly consumed by
star formation for the model with larger  $C_{\rm SF}$ 
\label{fig-1}}

\figcaption{Dependence of nonhomology parameters, 
$k_{\rm S}$,  $k_{\rm K}$,  and  $k_{\rm M}$
on $C_{\rm SF}$ for a set of models, Model 1.
Physical meanings of the three nonhomology parameters
are described in the manuscript.
The parameter values projected onto $xy$, $xz$, 
and $yz$ plane are plotted
by open squares for five models with different $C_{\rm SF}$ 
in the Model 1.
A solid line in each panel indicates the best fitted line derived
by least square fitting procedure for each set of results.
\label{fig-2}}

\figcaption{Dependence of  
${k_{\rm K}}^{-1}$  and  $k_{\rm M}$
on $k_{\rm S}$ for a set of models, Model 1.
The parameter values projected onto $xy$, $xz$, 
and $yz$ plane are plotted
by open squares for five models with different $C_{\rm SF}$ 
in the Model 1.
A solid line in each panel indicates the best fitted line derived
by least square fitting procedure for each set of results.
\label{fig-3}}

\figcaption{Dependence of $K_{\rm FP}$ 
on $C_{\rm SF}$ for a set of models, Model 1.
Physical meanings of the $K_{\rm FP}$
is described in the manuscript.
The parameter values of the  $K_{\rm FP}$
projected onto $xy$, $xz$, 
and $yz$ plane are  plotted
by open squares for models with different $C_{\rm SF}$ 
in the Model 1.
A solid line in each panel indicates the best fitted line derived
by least square fitting procedure for each set of results.
\label{fig-4}}

\figcaption{Dependence of $k_{\rm S}$ (top),
$k_{\rm K}$ (the second from the top), $k_{\rm M}$ (the third from the top),
and $K_{\rm FP}$ (bottom) on the $C_{\rm SF}$ for models 
in Group  A. 
Each 
open square represents the results for  each  of three projections
($xy$, $xz$  and $yz$ plane) in each model. 
A solid line in each panel indicates the best fitted line derived
by least square fitting procedure for each set of results.
\label{fig-5}}

\figcaption{The same as Fig. 5 but for models in  Group  B. 
\label{fig-6}}

\figcaption{The same as Fig. 5 but for models in  Group  C.
\label{fig-7}}

\figcaption{The same as Fig. 5 but for models in  Group  D. 
\label{fig-8}}

\figcaption{The same as Fig. 5 but for models in  Group  E. 
\label{fig-9}}

\figcaption{The Fundamental Plane (FP) in the present dissipative merger model.
The way of constructing the FP is described in the manuscript.
The solid line indicates  the best fitted line with the same slope
as that observed in the $K$ band FP (Pahre et al. 1995; Djorgovski et al. 1996).
Each open square represents the results for  each  of three projections 
($xy$, $xz$  and $yz$ plane) in each model.
\label{fig-10}}


\end{document}